\documentclass[onecolumn]{IEEEtran}
\usepackage{amssymb,amsmath,amsfonts,amsthm}
\usepackage[font=small,labelfont=bf]{caption}
\usepackage{color}
\usepackage{cases}
\usepackage{mathtools}
\usepackage{float}
\usepackage{cite}
\usepackage{suffix}
\usepackage{dblfloatfix}
\restylefloat{table}
\usepackage{amsfonts}
\usepackage{booktabs}
\usepackage{siunitx}
\usepackage{mathrsfs}
\usepackage{relsize}
\usepackage{graphicx}	
\usepackage{algorithm}
\newlength\figureheight 
\newlength\figurewidth
\usepackage{tabularx}
\usepackage{enumerate}
\usepackage{mathrsfs}
\usepackage{amsbsy}
\usepackage{units}
\usepackage{bm}

\usepackage{textcomp}
\usepackage{placeins}
\usepackage{bbm}
\newcommand{\ROME}[1]{%
\textup{\uppercase\expandafter{\romannumeral#1}}%
}
\usepackage{setspace}
\DeclarePairedDelimiterX\MeijerM[3]{\lparen}{\rparen}%
{\begin{smallmatrix}#1 \\ #2\end{smallmatrix}\delimsize\vert\,#3}
\newcommand\MeijerG[8][]{%
\mathcal{G}^{\,#2,#3}_{#4,#5}\MeijerM[#1]{#6}{#7}{#8}}
\WithSuffix\newcommand\MeijerG*[7]{%
\mathcal{G}^{\,#1,#2}_{#3,#4}\MeijerM*{#5}{#6}{#7}}
\newtheorem*{thm*}{Theorem}

\DeclareMathSizes{10}{8}{7}{7}
\usepackage{letltxmacro}

\makeatletter
\let\MYcaption\@makecaption
\makeatother
\usepackage[font=small]{subcaption}
\makeatletter
\let\@makecaption\MYcaption
\makeatother

\allowdisplaybreaks
  
\begin{document}

\title{{\huge Self-Interference in Full-Duplex Multi-User MIMO Channels}} 

\author{Arman Shojaeifard, \textit{Member,~IEEE}, Kai-Kit Wong, \textit{Fellow,~IEEE}, Marco Di Renzo, \textit{Senior Member,~IEEE},\\ Gan Zheng, \textit{Senior Member,~IEEE}, Khairi Ashour Hamdi, \textit{Senior Member,~IEEE}, Jie Tang, \textit{Member,~IEEE}
\thanks{A. Shojaeifard and K.-K Wong are with the Department of Electronic and Electrical Engineering, University College London, London, United Kingdom (e-mail: a.shojaeifard@ucl.ac.uk; kai-kit.wong@ucl.ac.uk).\par M. Di Renzo is with the Laboratoire des Signaux et Syst\`emes, CNRS, CentraleSup\'elec, Univ Paris Sud, Universit\'e Paris-Saclay, Gif-sur-Yvette, France (e-mail: marco.direnzo@l2s.centralesupelec.fr).\par G. Zheng is with the Wolfson School of Mechanical, Electrical and Manufacturing Engineering, Loughborough University, Loughborough, United Kingdom (e-mail: g.zheng@lboro.ac.uk).\par K. A. Hamdi is with the School of Electrical and Electronic Engineering, University of Manchester, Manchester, United Kingdom (e-mail: k.hamdi@ manchester.ac.uk).\par J. Tang is with the School of Electronic and Information Engineering, South China University of Technology, Guangzhou, China (e-mail: eejtang@scut.edu.cn).\par This work was supported by the Engineering and Physical Sciences Research Council (EPSRC) under grants EP/N008219/1 and EP/N007840/1.}}
\maketitle

\begin{abstract}
 
We consider a multi-user multiple-input multiple-output (MIMO) setup where full-duplex (FD) multi-antenna nodes apply linear beamformers to simultaneously transmit and receive multiple streams over Rician fading channels. The exact first and second positive moments of the residual self-interference (SI), involving the squared norm of a sum of non-identically distributed random variables, are derived in closed-form. The method of moments is hence invoked to provide a Gamma approximation for the residual SI distribution. The proposed theorem holds under arbitrary linear precoder/decoder design, number of antennas and streams, and SI cancellation capability. 

\end{abstract}


\section{Introduction}

To date, wireless systems, have been designed under a separation of the transmit/receive functions, a.k.a., half-duplex (HD) mode. This is typically achieved via orthogonal radio frequency (RF) partitioning, e.g., in time-division duplex (TDD) and frequency-division duplex (FDD) systems. The motivation behind this trend has been to avoid the overwhelming self-interference (SI) arising from the bi-directional operation. 

The rapid increase in traffic, under limited RF resources, however, has driven the incentive for an overhaul in the wireless system design. A candidate solution involves transceiving information over the same RF resources, i.e., full-duplex (FD) mode \cite{6353396}. There is a growing surge of interest in this topic, as in theory, FD has the potential to double the spectral efficiency compared to HD. Several point-to-point prototypes of FD radio nodes have been demonstrated in practice recently \cite{179789}. 

Despite pioneering efforts, FD operation in the context of multi-user multiple-input multiple-output (MIMO) setups is not well understood. A long-standing limitation is in the modeling of the residual SI using the Rayleigh distribution \cite{6971223}. This approach, whilst improving on the common perfect SI cancellation assumption, differs from measurements showing the residual SI channel undergoes Rician fading \cite{6353396}. The residual SI distribution over Rician fading channels is thus required for facilitating performance analysis and optimization.  

We consider a FD multi-antenna node communicating with multiple FD radios over the same RF resources. The Rician fading model is employed to capture the residual SI under arbitrary cancellation through tuning of the distribution parameters by design or measurements. With linear processing, it is not feasible to derive the statistics of the residual SI channel power gain directly. Here, exact closed-form expressions for the residual SI first and second positive moments are developed. We then exploit the method of moments in order to obtain an explicit Gamma approximation for the distribution of the residual SI over FD multi-user MIMO Rician fading channels. The validity of the theoretical findings is confirmed through simulations of the entire RF transmit/receive processing chain.

\textit{Notation:} $\boldsymbol{X}$ is a matrix with ($n,m$)-th entry $\{ \boldsymbol{X} \}_{n,m}$; $\boldsymbol{x}$ is a vector with $k$-th element $\{ \boldsymbol{x} \}_{k}$; $T$, $\dag$, and $+$ are the transpose, Hermitian-transpose, and pseudo-inverse; $\mathbb{E}\{.\}$ is the expected value; $\mathbb{V} \{ . \}$ is the variance; $\mathcal{P}(.)$ is the probability density function (pdf); $| . |$ is the modulus; $\| . \|$ is the norm; and $I_{0}(.)$ is the zeroth-order Bessel function of the first kind, respectively.

\section{System Model}

Consider $L$ cells, where in each cell $l$, $l \in \mathcal{L} = \{1,...,L\}$, a FD multi-antenna node $l_{0}$ communicates with respect to multiple FD radios $l_{k}$, $k \in \mathcal{K} = \{ 1,...,K \}$. Let $M$ and $N$ respectively denote the number of transmit and receive antennas at the FD multi-antenna nodes. The FD radios have two-antennas, one for transmission, and the other for reception. An application example is FD multi-antenna base stations communicating in the downlink/uplink with multiple FD mobile terminals. Assuming $K \leq \min(N, M)$, channel assignment is not required.     

Let $\boldsymbol{h}_{[l_{k},j_{0}]} \in \mathcal{C}^{1 \times M}$ and $\boldsymbol{h}_{[j_{0},l_{k}]} \in \mathcal{C}^{N \times 1}$ respectively denote the transmit and receive channels between the FD multi-antenna node in cell $j$ and the $k$-th FD radio in cell $l$. The respective combined channels are $\boldsymbol{H}_{[l_{K},j_{0}]} = [\boldsymbol{h}^{T}_{[l_{k},j_{0}]} ]^{T}_{1 \leq k \leq K} \in \mathcal{C}^{K \times M}$ and $\boldsymbol{H}_{[j_{0},l_{K}]} = [\boldsymbol{h}_{[j_{0},l_{k}]}]_{1 \leq k \leq K} \in \mathcal{C}^{N \times K}$. The cross-mode channels between $l_{0}$ and $j_{0}$, and between $l_{k}$ and $j_{k}$, are $\boldsymbol{H}_{[j_{0},l_{0}]} \in \mathcal{C}^{N \times M}$, and $h_{[j_{k},l_{k}]}$, respectively. The residual SI channels are Rician distributed with elements drawn from the complex Gaussian distribution $\mathcal{C} \mathcal{N} ( \mu , \nu^2 )$ \cite{6847175}, \cite{atzeni2015full}. Other channels are subject to Rayleigh fading with elements drawn from $\mathcal{C} \mathcal{N} \left( 0 , 1 \right)$. Perfect channel state information (CSI) is assumed.

We consider arbitrary linear beamforming design at the FD multi-antenna nodes. Let $\boldsymbol{s}_{[l_{K},l_{0}]} = [s_{[l_{k},l_{0}]}]^{T}_{1 \leq k \leq K} \in \mathcal{C}^{K \times 1}$, $\mathbb{E} \{ | s_{[l_{k},l_{0}]} |^{2} \} = 1$, denote the complex information vector from $l_{0}$ to all $l_{k}$. The corresponding complex information vector in the reverse communications direction is $\boldsymbol{s}_{[l_{0},l_{K}]} = [ s_{[l_{0},l_{k}]} ]^{T}_{1 \leq k \leq K} \in \mathcal{C}^{K \times 1}$, $\mathbb{E} \{ | s_{[l_{0},l_{k}]} |^2 \} = 1$. The transmit signal vector under linear precoding at $l_{0}$ is hence constructed as $\boldsymbol{t}_{[l_{K},l_{0}]} = \boldsymbol{V}_{[l_{K},l_{0}]} \boldsymbol{s}_{[l_{K},l_{0}]}$ where $\boldsymbol{V}_{[l_{K},l_{0}]} = [\boldsymbol{v}_{[l_{k},l_{0}]}]_{1 \leq k \leq K} \in \mathcal{C}^{M \times K}$ is the precoding matrix. In addition, the linear receive filter at $l_{0}$ is defined as $\boldsymbol{W}_{[l_{0},l_{K}]} = [ \boldsymbol{w}^{T}_{[l_{0},l_{k}]} ]^{T}_{1 \leq k \leq K} \in \mathcal{C}^{K \times N}$. 

The received baseband signal from the FD multi-antenna node $l_{0}$ at its FD radio $l_{k}$ is expressed as
\begin{align}
y_{[l_{k},l_{0}]} & = \underbrace{ \boldsymbol{h}_{[l_{k},l_{0}]} \boldsymbol{v}_{[l_{k},l_{0}]} s_{[l_{k},l_{0}]}}_{\text{useful signal}, x_{[l_{k},l_{0}]}} + \underbrace{ \boldsymbol{h}_{[l_{k},l_{0}]} \sum_{u \in \mathcal{K} \backslash \{ k \}} \boldsymbol{v}_{[l_{u},l_{0}]} s_{[l_{u},l_{0}]}}_{\text{multi-user interference}, mui_{[l_{k},l_{0}]}} + \underbrace{\sum_{j \in \mathcal{L} \backslash \{ l \}} \boldsymbol{h}_{[l_{k},j_{0}]} \boldsymbol{V}_{[j_{K},j_{0}]} \boldsymbol{s}_{[j_{K},j_{0}]} }_{\text{inter-cell interference}, ici_{[l_{k},l_{0}]}} \nonumber \\ & + \underbrace{\sum_{j \in \mathcal{L}, u \in \mathcal{K} \backslash \{ l,k \}} h_{[l_{k},j_{u}]} s_{[j_{0},j_{u}]} }_{\text{cross-mode interference}, \, cmi_{[l_{k},l_{0}]}} + \underbrace{ h_{[l_{k},l_{k}]} s_{[l_{0},l_{k}]} }_{\text{residual self-interference}, si_{[l_{k},l_{0}]}} + \underbrace{\eta_{[l_{k},l_{0}]}}_{\text{noise}, n_{[l_{k},l_{0}]}}
\label{eq:baseband-d}
\end{align}
where $\eta_{[l_{k},l_{0}]}$ is the complex additive white Gaussian noise (AWGN). The post-processing received baseband signal in the reverse communications direction is written as
\begin{align}
y_{[l_{0},l_{k}]} & = \underbrace{ \boldsymbol{w}^{T}_{[l_{0},l_{k}]}  \boldsymbol{h}_{[l_{0},l_{k}]} s_{[l_{0},l_{k}]}}_{\text{useful signal}, x_{[l_{0},l_{k}]}} + \underbrace{ \boldsymbol{w}^{T}_{[l_{0},l_{k}]} \sum_{u \in \mathcal{K} \backslash \{ k \}} \boldsymbol{h}_{[l_{0},l_{u}]} s_{[l_{0},l_{u}]} }_{\text{multi-user interference}, mui_{[l_{0},l_{k}]}} + \underbrace{ \boldsymbol{w}^{T}_{[l_{0},l_{k}]} \sum_{j \in \mathcal{L},u \in \mathcal{K} \backslash \{ l , k \}} \boldsymbol{h}_{[l_{0},j_{u}]} s_{[j_{0},j_{u}]} }_{\text{inter-cell interference}, ici_{[l_{0},l_{k}]}} \nonumber \\ & + \underbrace{ \boldsymbol{w}^{T}_{[l_{0},l_{k}]} \sum_{j \in \mathcal{L} \backslash \{ l \}}  \boldsymbol{H}_{[l_{0},j_{0}]} \boldsymbol{V}_{[j_{K},j_{0}]} \boldsymbol{s}_{[j_{K},j_{0}]} }_{\text{cross-mode interference}, \, cmi_{[l_{0},l_{k}]}} + \underbrace{ \boldsymbol{w}^{T}_{[l_{0},l_{k}]}  \boldsymbol{H}_{[l_{0},l_{0}]} \boldsymbol{V}_{[l_{K},l_{0}]} \boldsymbol{s}_{[l_{K},l_{0}]} }_{\text{residual self-interference}, si_{[l_{0},l_{k}]}} + \underbrace{\boldsymbol{w}^{T}_{[l_{0},l_{k}]} \boldsymbol{\eta}_{[l_{0},l_{K}]}}_{\text{scaled noise}, n_{[l_{0},l_{k}]}}
\label{eq:baseband-u}
\end{align}
where $\boldsymbol{\eta}_{[l_{0},l_{K}]} \in \mathcal{C}^{N \times 1}$ is the circularly-symmetric complex AWGN vector.

\section{Signals Statistics}

Next, we formulate the signal-to-interference-plus-noise ratios (SINRs). Note $d$ and $u$ are respectively used in place of $[l_{k},l_{0}]$ and $[l_{0},l_{k}]$ where the context is clear. Hence,
\begin{align}
\mathcal{Y}_{y_{d}} = \frac{\mathcal{X}_{x_{d}}}{\mathcal{I}_{mui_{d}} + \mathcal{I}_{ici_{d}} + \mathcal{I}_{cmi_{d}} + \mathcal{I}_{si_{d}} + \mathcal{N}_{n_{d}}}
\end{align}
where we have $\mathcal{X}_{x_{d}} = | \boldsymbol{h}_{[l_{k},l_{0}]} \boldsymbol{v}_{[l_{k},l_{0}]} |^2$, $\mathcal{I}_{mui_{d}} = \sum_{u \in \mathcal{K} \backslash \{ k \}}$ $| \boldsymbol{h}_{[l_{k},l_{0}]} \boldsymbol{v}_{[l_{u},l_{0}]} |^2$, $\mathcal{I}_{ici_{d}} = 
\sum_{j \in \mathcal{L} \backslash \{ l \} } \| \boldsymbol{h}_{[l_{k},j_{0}]} \boldsymbol{V}_{[j_{K},j_{0}]} \|^2$, $\mathcal{I}_{cmi_{d}} =  \sum_{j \in \mathcal{L} , u \in \mathcal{K} \backslash \{ l,k \}} | h_{[l_{k},j_{u}]} |^2$, $\mathcal{I}_{si_{d}} =  | h_{[l_{k},l_{k}]} |^2$, and $\mathcal{N}_{n_{d}} = | \eta_{[l_{k},l_{0}]} |^{2}$ respectively. Moreover, 
\begin{align}
\mathcal{Y}_{y_{u}} = \frac{\mathcal{X}_{x_{u}}}{\mathcal{I}_{mui_{u}} + \mathcal{I}_{ici_{u}} + \mathcal{I}_{cmi_{u}} + \mathcal{I}_{si_{u}} + \mathcal{N}_{n_{u}}}
\end{align}
with the signals $\mathcal{X}_{x_{u}} =  | \boldsymbol{w}^{T}_{[l_{0},l_{k}]} \boldsymbol{h}_{[l_{0},l_{k}]} |^2$,
$\mathcal{I}_{mui_{u}} = \sum_{u \in \mathcal{K} \backslash \{ k \}}$ $| \boldsymbol{w}^{T}_{[l_{0},l_{k}]} \boldsymbol{h}_{[l_{0},l_{u}]} |^2$, $\mathcal{I}_{ici_{u}} = \sum_{j \in \mathcal{L}, u \in \mathcal{K} \backslash \{ l,k \}} | \boldsymbol{w}^{T}_{[l_{0},l_{k}]} \boldsymbol{h}_{[l_{0},j_{u}]} |^{2}$, $\mathcal{I}_{cmi_{u}} = \sum_{j \in \mathcal{L} \backslash \{ l \}} \| \boldsymbol{w}^{T}_{[l_{0},l_{k}]} \boldsymbol{H}_{[l_{0},j_{0}]} \boldsymbol{V}_{[j_{K},j_{0}]} \|^{2}$, $\mathcal{I}_{si_{u}} =  \| \boldsymbol{w}^{T}_{[l_{0},l_{k}]}$ $\boldsymbol{H}_{[l_{0},l_{0}]} \boldsymbol{V}_{[l_{K},l_{0}]} \|^{2}$, and $\mathcal{N}_{n_{u}} = | \boldsymbol{w}^{T}_{[l_{0},l_{k}]} \boldsymbol{\eta}_{[l_{0},l_{K}]} |^{2}$.

With linear processing over isotropic MIMO Rayleigh fading channels, the signals can be captured using the Gamma distribution \cite{7478073}. 
E.g., consider linear zero-forcing (ZF) for removing multi-user interference by setting (i) the column vectors of $\boldsymbol{V}_{[l_{K},l_{0}]}$ equal to the normalized columns of $\boldsymbol{H}^{+}_{[l_{K},l_{0}]} = \boldsymbol{H}^{\dag}_{[l_{K},l_{0}]} ( \boldsymbol{H}_{[l_{K},l_{0}]} \boldsymbol{H}^{\dag}_{[l_{K},l_{0}]} )^{-1}$, and (ii) the row vectors of $\boldsymbol{W}_{[l_{0},l_{K}]}$ equal to the normalized rows of $\boldsymbol{H}^{+}_{[l_{0},l_{K}]} = ( \boldsymbol{H}^{\dag}_{[l_{0},l_{K}]} \boldsymbol{H}_{[l_{0},l_{K}]} )^{-1} \boldsymbol{H}^{\dag}_{[l_{0},l_{K}]}$. With the projection of each useful channel vector onto the nullspace spanned by the multi-user interference, we have $| \boldsymbol{h}_{[l_{k},l_{0}]} \boldsymbol{v}_{[l_{k},l_{0}]} |^2 \sim \text{Gamma} (M - K + 1,1) $ and $| \boldsymbol{w}^{T}_{[l_{0},l_{k}]} \boldsymbol{h}_{[l_{0},l_{k}]} |^2 \sim \text{Gamma} (N - K + 1,1)$. Moreover, $| \boldsymbol{w}^{T}_{[l_{0},l_{k}]} \boldsymbol{h}_{[l_{0},j_{u}]} |^{2} \sim \text{Gamma} (1,1)$ and $| h_{[l_{k},j_{u}]} |^2 \sim \text{Gamma} (1 ,1)$. Further, under the assumption that the outer-cell precoding vectors are independent, we have $\| \boldsymbol{h}_{[l_{k},j_{0}]} \boldsymbol{V}_{[j_{K},j_{0}]} \|^2 \sim \text{Gamma} (U,1)$ and $\| \boldsymbol{w}^{T}_{[l_{0},l_{k}]} \boldsymbol{H}_{[l_{0},j_{0}]} \boldsymbol{V}_{[j_{K},j_{0}]} \|^{2} \sim \text{Gamma} (U,1)$. 

\begin{table*}[!t] 
\centering 
\renewcommand{\arraystretch}{1.5}
\begin{tabular}{|c| c| c|}
\hline
{\small Single-User} & $\kappa = \tfrac{(N + 1)(M + 1) \big(\mu^2 + \nu^2 \big)^2}{ (3 N M - N - M - 1) \mu^4  + 2 (N + 1) (M + 1) \mu^2 \nu^2 + (N + 1) (M + 1) \nu^4}$ & $\theta = \mu ^2+\nu ^2+\tfrac{2 (M N - N - M - 1) \mu^4}{(N+1) (M+1) (\mu^2 + \nu^2)}$ \\  
\hline 
{\small Rayleigh Channel} & $\kappa = \tfrac{K (\max (N,M) - K + 2)}{\max (N,M) + 1} $ & $\theta = \tfrac{\max (N,M) + 1}{\max (N,M) - K + 2}$ \\  
\hline 
{\small Massive MIMO} & $\kappa = \tfrac{K \big( \mu^2 + \nu^2 \big)^2}{(K+2) \mu^4 + 2 \mu^2 \nu^2 + \nu^4}$ & $\theta = \tfrac{(K+2) \mu^4 + 2 \mu^2 \nu^2 + \nu^4}{\mu^2 + \nu^2}$ \\ 
\hline 
\end{tabular}
\caption*{{\small \textbf{Table.} Residual SI distribution using Gamma moment matching in some special cases of interest.}}
\label{tbl:SpecialCases}
\end{table*}

Next, we consider the residual SI channel of an arbitrary FD radio $l_{k}$, with single transmit and receive antennas, $h_{[l_{k},l_{0}]} \sim \mathcal{C} \mathcal{N} (\mu,\nu^2)$. The residual SI channel power gain, $\mathcal{I}_{si_{d}} = | h_{[l_{k},l_{0}]} |^{2}$, has a non-central Chi-squared distribution 
\begin{align}
\mathcal{P}_{\mathcal{I}_{si_{d}}} (x) & = \tfrac{1 + \varpi}{\Omega} \exp \left( - \left( \varpi + \tfrac{(1 + \varpi) x}{\Omega} \right) \right) I_{0} \left(2 \sqrt{ \tfrac{\varpi (1 + \varpi) x}{\Omega} }\right)
\label{eq:SI-SISO}
\end{align}
where $\varpi$ and $\Omega$ are the Rician factor and fading attenuation with $\mu \triangleq \sqrt{\frac{\varpi \Omega}{\varpi + 1}}$ and $\nu \triangleq \sqrt{\frac{\Omega}{\varpi+1}}$, respectively. The method of moments can be applied to provide a Gamma approximation for the residual SI in (\ref{eq:SI-SISO}). Hence, the residual SI channel power gain of an arbitrary FD radio $k$, with single transmit and receive antennas, and fading coefficients drawn from $\mathcal{C} \mathcal{N} (\mu,\nu^2)$, is approximated using $\mathcal{I}_{si_{d}} = | h_{[l_{k},l_{0}]} |^{2} \sim \text{Gamma} \left( \kappa,\theta \right)$ where
\begin{align}
\kappa \triangleq \tfrac{ \big( \mu^2+\nu^2 \big)^2}{\left( 2 \mu^2 + \nu^2 \right) \nu^2 } 
\end{align}
and
\begin{align}
\theta \triangleq \tfrac{\big( 2 \mu^2 + \nu^2 \big) \nu^2 }{\mu^2+\nu^2}.
\end{align}

The residual SI channel power gain at the FD multi-antenna nodes, under arbitrary number of antennas and information streams, involves the squared norm of a sum of non-identically distributed random variables. For instance, for the decoding of the transmitted signal from the arbitrary FD node $l_{k}$ at the receiver of the FD multi-antenna node $l_{0}$, we have $\mathcal{I}_{si_{u}} = \| \boldsymbol{w}^{T}_{[l_{0},l_{k}]} \boldsymbol{H}_{[l_{0},l_{0}]} \boldsymbol{V}_{[l_{K},l_{0}]} \|^{2}$ where $\boldsymbol{w}^{T}_{[l_{0},l_{k}]} \boldsymbol{H}_{[l_{0},l_{0}]} \boldsymbol{V}_{[l_{K},l_{0}]} = [ \sum^{N}_{n = 1} \sum^{M}_{m = 1} \{ \boldsymbol{w}^{T}_{[l_{0},l_{k}]} \}_{n} \{ \boldsymbol{H}_{[l_{0},l_{0}]} \}_{n,m} \{ \boldsymbol{V}_{[l_{K},l_{0}]}  \}_{m,k} ]_{1 \leq k \leq K}$.

It is therefore not possible to directly derive the exact pdf of $\mathcal{I}_{si_{u}}$. As a result, we derive closed-form expressions for the exact first and second positive moments of $\mathcal{I}_{si_{u}}$. We then apply the method of moments in order to develop a unified closed-form approximation for the residual SI distribution over multi-user MIMO Rician fading channels.

\begin{thm*} 
\label{thm:SI-MIMO}
Consider the residual SI fading channel of an arbitrary FD multi-antenna node $l_{0}$ with $M$ transmit and $N$ receive antennas, $\boldsymbol{H}_{[l_{0},l_{0}]}$, comprising elements distributed according to $\mathcal{C}\mathcal{N} (\mu,\nu^{2})$. Using a linear precoder $\boldsymbol{V}_{[l_{K},l_{0}]}$ and a linear decoder $\boldsymbol{W}_{[l_{0},l_{K}]}$, $l_{0}$ simultaneously transmits and receives information streams with respect to $K$ FD radios, each equipped with single transmit and receive antennas. The corresponding residual SI channel power gain at the FD multi-antenna node $l_{0}$ can be approximated using $\mathcal{I}_{si_{u}} = \| \boldsymbol{w}^{T}_{[l_{0},l_{k}]} \boldsymbol{H}_{[l_{0},l_{0}]} \boldsymbol{V}_{[l_{K},l_{0}]} \|^{2} \sim \text{Gamma} \left( \kappa,\theta \right)$ where
\begin{align}
\kappa 
\triangleq  \tfrac{ K (N + 1) \left( M - K + 2 \right) \left( \mu^{2} + \nu^{2} \right)^{2}}{\left( 2 N M + \tfrac{K \left( M - K + 2 \right) }{(M + 1)} ( N M - N - M - 1) \right) \mu^4 + (N + 1) (M + 1) \nu ^2 \left( 2 \mu ^2 + \nu^2 \right)}   
\end{align}
and
\begin{align}
\theta 
\triangleq \tfrac{\left( 2 N M + \tfrac{K \left( M - K + 2 \right) }{(M + 1)} ( N M - N - M - 1) \right) \mu^4 + (N + 1) (M + 1) \nu ^2 \left( 2 \mu ^2 + \nu^2 \right)}{ (N + 1) \left( M - K + 2 \right) \left( \mu^{2} + \nu^{2} \right)}. 
\end{align}
\textit{Proof:} See Appendix.
\end{thm*}

The unified expression in the \textit{Theorem} holds for arbitrary linear precoder/decoder design, number of transmit/receive antennas, number of streams, and Rician channel statistics (and in turn arbitrary SI cancellation capability). The distribution of the residual SI in special scenarios of interest can also be readily obtained in closed-form. Some useful examples are provided for the sake of demonstration in the \textit{Table}.       

\section{Performance Analysis}

\begin{figure*}
     \begin{minipage}[b]{0.5\textwidth}
          \centering
          \includegraphics{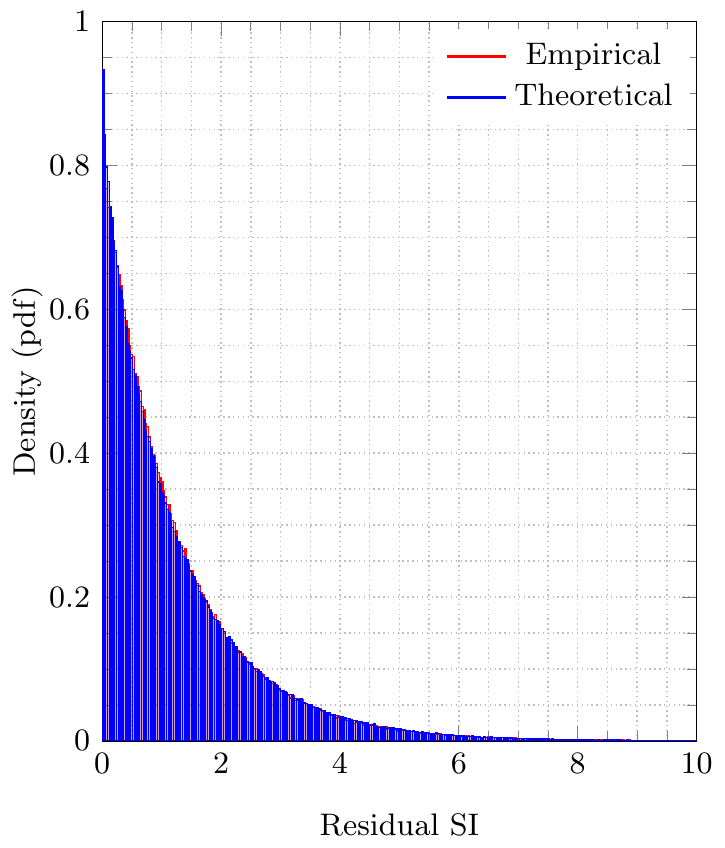}
     \end{minipage}%
     \begin{minipage}[b]{0.5\textwidth}
          \centering
	      \includegraphics{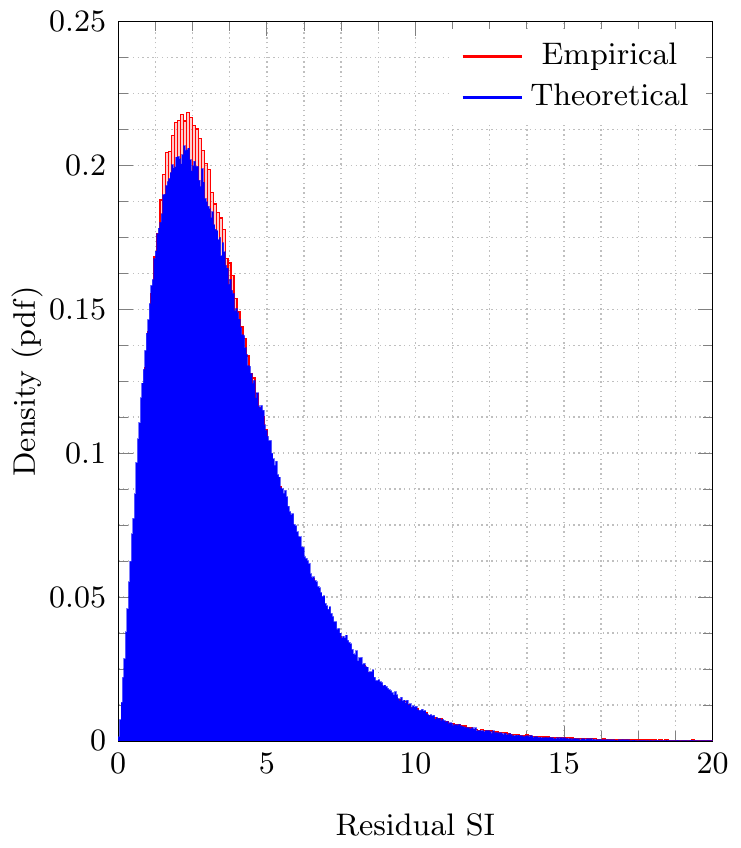}   		          
     \end{minipage}%
\caption*{{\small \textbf{Figure.} Simulation parameters are: $M = 16$, $N = 8$, $\mu = \tfrac{1}{2}$, $\nu = 1$, $\text{MC} = 10^{7}$, $K = 1$ (left), $K = 3$ (right).}}
 \end{figure*}

Here, we perform simulations of the entire RF transmit/receive processing chain in order to assess the validity of the theoretical findings. The step-by-step sketches of the different simulation methodologies are described below. 

\paragraph*{Empirical approach}
1) Select the parameters $M$, $N$, $K$, $\mu$, and $\nu$. 2) Generate channel matrices $\boldsymbol{H}_{[l_{0},l_{0}]}$, using $\mathcal{C} \mathcal{N} (\mu,\nu^{2})$, and $\boldsymbol{H}_{[l_{K},l_{0}]}$, $\boldsymbol{H}_{[l_{0},l_{k}]}$, using $\mathcal{C} \mathcal{N} (0,1)$. 3) Generate complex information vectors $\boldsymbol{s}_{[l_{K},l_{0}]}$ and $\boldsymbol{s}_{[l_{0},l_{K}]}$. 4) Design linear beamformers $\boldsymbol{V}_{[l_{K},l_{0}]}$ and $\boldsymbol{W}_{[l_{0},l_{K}]}$ using CSI. 5) Compute the corresponding $\mathcal{I}_{si_{d}}$. 6) Repeat the above process for $\text{MC}$ trials. 7) Plot the empirical residual SI distribution.

\paragraph*{Theoretical approach}
1-2) Same as above. 3) Generate $\mathcal{I}_{si_{d}}$ from the Gamma distribution in the \textit{Theorem}. 4) Repeat for $\text{MC}$ trials. 5) Plot the theoretical residual SI \nolinebreak[4] distribution. 

Without loss of generality, consider a $16 \times 8$ MIMO Rician fading channel with mean $\tfrac{1}{2}$ and variance $1$. The distributions of the residual SI with linear ZF precoding and decoding are depicted with different number of information streams in the \textit{Figure}. It can be observed that the theoretical data provides a near exact fit for single-user MIMO (left figure) and a tight approximation for multi-user MIMO (right figure) with respect to the empirical data, respectively. Note that the moment matching accuracy increases for smaller $\mu$, $\nu$, and $K$. Increasing $N$ and $M$, for example in the context of massive MIMO, on the other hand, enhances the goodness of the fit.         

\section{Summary}

A rigorous study of the residual SI over FD multi-user MIMO Rician fading channels was provided. We considered FD multi-antenna nodes applying linear beamformers to communicate with multiple FD radios. The residual SI fading channels were drawn from a complex Gaussian distribution with arbitrary statistics. We derived the exact first and second positive moments of the residual SI in closed-form. The Gamma moment matching approximation was then adopted to develop a unified expression for the residual SI distribution. 

\appendices

\bibliographystyle{IEEEtran}
\bibliography{IEEEabrv,myref}

\section*{Appendix}
\label{app:SI-MIMO}

Under Rician fading with elements drawn from $\mathcal{C} \mathcal{N} \big( \mu,\nu^{2} \big)$, $\mathbb{E} \big\{ | \{ \boldsymbol{H}_{[l_{0},l_{0}]} \}_{n,m} |^{2} \big\} = \mu ^2+\nu ^2$, $\mathbb{E} \big\{ | \{ \boldsymbol{H}_{[l_{0},l_{0}]} \}_{n,m} |^4 \big\}$ $= \mu^{2} \big( \mu^2 + 4 \nu^2 \big) + 2 \nu^4$, and $\mathbb{V} \{ | \{ \boldsymbol{H}_{[l_{0},l_{0}]} \}_{n,m} |^{2} \} = \nu^{2} \big( 2 \mu^2 + \nu^2 \big)$. Considering arbitrary linear precoder/decoder design, we can derive $\mathbb{E} \big\{ | \{ \boldsymbol{V}_{[l_{K},l_{0}]}  \}_{m,k} |^{2} \big\} = \tfrac{1}{M}$, $\mathbb{E} \big\{ | \{ \boldsymbol{V}_{[l_{K},l_{0}]}  \}_{m,k} |^4 \big\} = \tfrac{2}{M ( M + 1 )}$, $\mathbb{V} \big\{ | \{ \boldsymbol{V}_{[l_{K},l_{0}]}  \}_{m,k} |^{2} \big\} = \tfrac{M - 1}{M^{2} (Nt + 1)}$, $\mathbb{E} \big\{ | \{ \boldsymbol{w}^{T}_{[l_{0},l_{k}]} \}_{n} |^{2} \big\} =\tfrac{1}{N}$,\linebreak $\mathbb{E} \big\{ | \{ \boldsymbol{w}^{T}_{[l_{0},l_{k}]} \}_{n} |^4 \big\} = \tfrac{2}{N (N + 1)}$, and $\mathbb{V} \big\{ | \{ \boldsymbol{w}^{T}_{[l_{0},l_{k}]} \}_{n} |^{2} \big\} = \tfrac{N - 1}{N^{2} (N + 1)}$. We use $\boldsymbol{\delta}_{[l_{0},l_{0}]} \in \mathcal{C}^{\min (N,M) \times 1}$ to denote the vector containing the $\min (N,M)$ largest singular-values of the residual SI channel at $l_{0}$. 

The corresponding first and second positive moments of the residual SI can be respectively developed as
\begin{multline}
\mathbb{E} \left\{ \| \boldsymbol{w}^{T}_{[l_{0},l_{k}]} \boldsymbol{H}_{[l_{0},l_{0}]} \boldsymbol{V}_{[l_{K},l_{0}]}  \|^{2} \right\} = \mathbb{E} \left\{ \sum^{K}_{k=1} |  \boldsymbol{w}^{T}_{[l_{0},l_{k}]} \boldsymbol{H}_{[l_{0},l_{0}]} \boldsymbol{v}_{[l_{k},l_{0}]} |^{2} \right\} \\ = \mathbb{E} \Biggr\{ \sum^{K}_{k = 1} \sum^{N}_{n_{1} = 1} \sum^{M}_{m_{1} = 1} \sum^{N}_{n_{2} = 1} \sum^{M}_{m_{2} = 1} \{ \boldsymbol{w}^{T}_{[l_{0},l_{k}]} \}_{n_{1}} \{ \boldsymbol{w}^{T}_{[l_{0},l_{k}]} \}^{\dag}_{n_{2}} \{ \boldsymbol{H}_{[l_{0},l_{0}]} \}_{n_{1},m_{1}} \{ \boldsymbol{H}_{[l_{0},l_{0}]} \}^{\dag}_{n_{2},m_{2}}  \{ \boldsymbol{V}_{[l_{K},l_{0}]}  \}_{m_{1},k} \{\boldsymbol{V}_{[l_{K},l_{0}]}  \}^{\dag}_{m_{2},k} \Biggr\} \\ = \mathbb{E} \left\{ \sum^{K}_{k = 1} \sum^{N}_{n = 1} \sum^{M}_{m = 1} | \{ \boldsymbol{w}^{T}_{[l_{0},l_{k}]} \}_{n} |^{2} | \{ \boldsymbol{H}_{[l_{0},l_{0}]} \}_{n,m} |^{2} | \{ \boldsymbol{V}_{[l_{K},l_{0}]}  \}_{m,k} |^{2} \right\}
= K ( \mu^2 + \nu^2 )
\label{1stmom}
\end{multline}
and
\begin{multline}
\mathbb{E} \left\{ \| \boldsymbol{w}^{T}_{[l_{0},l_{k}]} \boldsymbol{H}_{[l_{0},l_{0}]} \boldsymbol{V}_{[l_{K},l_{0}]}  \|^4 \right\} = \mathbb{E} \Biggr\{ \Big(  \sum^{K}_{k=1} |  \boldsymbol{w}^{T}_{[l_{0},l_{k}]} \boldsymbol{H}_{[l_{0},l_{0}]} \boldsymbol{v}_{[l_{k},l_{0}]} |^{2} \Big)^{2} \Biggr\} \\ = \mathbb{E} \Biggr\{ \sum^{K}_{k_{1} = 1} \sum^{N}_{n_{1} = 1} \sum^{N}_{n_{2} = 1} \sum^{M}_{m_{1} = 1} \sum^{M}_{m_{2} = 1} \sum^{K}_{k_{2} = 1} \sum^{N}_{n_{3} = 1} \sum^{N}_{n_{4} = 1} \sum^{M}_{m_{3} = 1} \sum^{M}_{m_{4} = 1} \{ \boldsymbol{w}^{T}_{[l_{0},l_{k}]} \}_{n_{1}} \{ \boldsymbol{w}^{T}_{[l_{0},l_{k}]} \}^{\dag}_{n_{2}}  \{ \boldsymbol{w}^{T}_{[l_{0},l_{k}]} \}_{n_{3}} \{ \boldsymbol{w}^{T}_{[l_{0},l_{k}]} \}^{\dag}_{n_{4}} \\ \times \{ \boldsymbol{H}_{[l_{0},l_{0}]} \}_{n_{1},m_{1}} \{ \boldsymbol{H}_{[l_{0},l_{0}]} \}^{\dag}_{n_{2},m_{2}}  \{ \boldsymbol{H}_{[l_{0},l_{0}]} \}_{n_{3},m_{3}} \{ \boldsymbol{H}_{[l_{0},l_{0}]} \}^{\dag}_{n_{4},m_{4}} \{ \boldsymbol{V}_{[l_{K},l_{0}]}  \}_{m_{1},k_{1}} \{\boldsymbol{V}_{[l_{K},l_{0}]}  \}^{\dag}_{m_{2},k_{1}} \{ \boldsymbol{V}_{[l_{K},l_{0}]}  \}_{m_{3},k_{2}} \{\boldsymbol{V}_{[l_{K},l_{0}]}  \}^{\dag}_{m_{4},k_{2}} \Biggr\} \\ = \mathbb{E} \left\{ \sum^{K}_{k = 1} \sum^{\min (N,M) }_{i = 1} | \{ \boldsymbol{w}^{T}_{[l_{0},l_{k}]} \}_{i} |^{4} | \{ \boldsymbol{\delta}_{[l_{0},l_{0}]} \}_{i} |^4 | \{\boldsymbol{V}_{[l_{K},l_{0}]} \}_{i,k} |^4 \right\} \\ + 2 \mathbb{E} \Biggr\{ \sum^{K}_{k = 1} \sum^{\min (N,M) }_{i_{1} = 1,i_{1} \neq i_{2}} \sum^{\min (N,M) }_{i_{2} = 1,i_{2} \neq i_{1}} | \{ \boldsymbol{w}^{T}_{[l_{0},l_{k}]} \}_{i_{1}} |^{2} \{ \boldsymbol{w}^{T}_{[l_{0},l_{k}]} \}_{i_{2}} |^{2} | \{ \boldsymbol{\delta}_{[l_{0},l_{0}]} \}_{i_{1}} |^2 | \{ \boldsymbol{\delta}_{[l_{0},l_{0}]} \}_{i_{2}} |^2 \{\boldsymbol{V}_{[l_{K},l_{0}]} \}_{i_{1},k} |^{2} | \{\boldsymbol{V}_{[l_{K},l_{0}]} \}_{i_{2},k} |^{2} \Biggr\} \\ + 2 \mathbb{E} \Biggr\{ \sum^{K - 1}_{k_{1} = 1} \sum^{K}_{k_{2} = k_{1} + 1}  \Big| \sum^{N}_{n = 1} \sum^{M}_{m = 1} \{ \boldsymbol{w}^{T}_{[l_{0},l_{k}]} \}_{n} \{ \boldsymbol{H}_{[l_{0},l_{0}]} \}_{n,m} \{ \boldsymbol{V}_{[l_{K},l_{0}]} \}_{m,k_{1}} \Big|^2 \Big| \sum^{N}_{n = 1} \sum^{M}_{m = 1} \{ \boldsymbol{w}^{T}_{[l_{0},l_{k}]} \}_{n} \{ \boldsymbol{H}_{[l_{0},l_{0}]} \}_{n,m} \{ \boldsymbol{V}_{[l_{K},l_{0}]} \}_{m,k_{2}} \Big|^2 \Biggr\} \\ = \tfrac{4 K}{(N+1) (M+1)} \left(N M \mu^4 + (N+M) \nu^2 \left( 2 \mu^2 + \nu^2 \right) \right) + \tfrac{2 \, K (N-1) (M-1)}{(N+1) (M+1)} \nu^2 \left(2 \mu^2 + \nu^2\right) \\ + K (K-1) \left(\tfrac{1}{M-K+2}+1\right) \left(\tfrac{2 N M}{N M + N + M +1} \mu ^4 + \nu^2 \left( 2 \mu^2 + \nu^2 \right) \right) = K \left(\tfrac{M+1}{M-K+2}+K\right) \left(\tfrac{2 N M}{N M+N+M+1} \mu^4 + \nu^2 \left( 2 \mu^2 + \nu^2 \right) \right).
\label{2ndmom}
\end{multline}
These moments can accordingly be used to derive an expression for the residual SI variance as\begin{multline}
\mathbb{V} \left\{ \| \boldsymbol{w}^{T}_{[l_{0},l_{k}]} \boldsymbol{H}_{[l_{0},l_{0}]} \boldsymbol{V}_{[l_{K},l_{0}]}  \|^2 \right\} = \mathbb{E} \left\{ \| \boldsymbol{w}^{T}_{[l_{0},l_{k}]} \boldsymbol{H}_{[l_{0},l_{0}]} \boldsymbol{V}_{[l_{K},l_{0}]}  \|^4 \right\} - \mathbb{E} \left\{ \| \boldsymbol{w}^{T}_{[l_{0},l_{k}]} \boldsymbol{H}_{[l_{0},l_{0}]} \boldsymbol{V}_{[l_{K},l_{0}]}  \|^{2} \right\}^{2} \\ = \tfrac{K}{M-K+2} \Biggr(\tfrac{K \left( M - K + 2 \right) ( N M - N - M - 1) + 2 N M (M+1)}{(N + 1) (M + 1)} \mu^4 + (M+1) \nu^2 \left( 2 \mu^2 + \nu^2 \right) \Biggr).
\label{fvar}
\end{multline}
Using (\ref{1stmom}) and (\ref{fvar}), we can apply the moment matching technique as in the \textit{Theorem} provided.
\hfill $\blacksquare$

\end{document}